\documentclass[aps,prd,reprint,nofootinbib]{revtex4-2}
\usepackage{amsmath,amssymb,bm}
\usepackage{hyperref}
\usepackage{mathtools}
\usepackage{physics}
\usepackage{graphicx}

\begin{document}

\title{Three-Body Holonomy as a Toy-Model Mechanism \\for Family Triplication and Mass Hierarchy in (1+1) Dimensions}

\author{Tadashi Yoshikawa}
\affiliation{Nagoya Aoi University, Nagoya, Japan}
\date{September 19, 2026}

\begin{abstract}
We investigate the phenomenological consequences of a genuine three-body holonomy and cyclic ordering structure
in a relativistic Dirac system in one spatial dimension. After removal of the center-of-mass coordinate, 
the triple-coincidence point punctures the two-dimensional relative configuration space and permits 
a nontrivial $U(1)$ winding phase. In contrast, the known pairwise Sakamoto--Munakata--Ino contact 
interactions produce a trivial net matching around the triple-coincidence point. 
The six particle-ordering sectors decompose, at fixed intrinsic parity, 
into a three-dimensional cyclic space. A Hermitian $C_3$-invariant effective mass operator 
on this space admits a Peierls-type realization in which the gauge-invariant phase accumulated 
around the three links equals the three-body holonomy $\theta_3$. Its eigenvalues are
\begin{equation*}
M^{[k]}=M_0+2\Delta\cos\!\left(\frac{\theta_3+2\pi k}{3}\right),\qquad k=0,1,2.
\end{equation*}
The holonomy lifts the conjugate-channel degeneracy and permits a parametrically light branch 
through cancellation between the common mass and the holonomy-induced shift. 
We then allow cyclic-symmetry breaking and consider a general Hermitian three-state mass matrix. 
Eliminating two heavy states by the exact Schur complement yields a low-energy correction 
containing the rephasing-invariant loop product $\mathrm{Re}(t_{12}t_{23}t_{31})\propto\cos\theta_3$. 
Thus, even when only one branch is kinematically accessible, 
its effective mass can retain a finite-memory contribution from the complete three-state loop. 
The complementary invariant $\mathrm{Im}(t_{12}t_{23}t_{31})\propto\sin\theta_3$ is phase sensitive 
but does not by itself establish CP violation; a measurable CP-odd observable requires 
additional noncommuting dynamics. The construction is therefore a low-dimensional phenomenological proof 
of concept for family-like triplication, mass hierarchy, and infrared memory, 
rather than a microscopic theory of Standard Model fermion generations.
\end{abstract}

\maketitle

\section{Introduction}

The replication of quarks and leptons into three generations is one of the basic empirical structures 
of the Standard Model (SM), but the SM does not determine why the number of families is three. 
The origin of family replication and of the associated fermion-mass and mixing hierarchies 
remains part of the broader flavor problem. Proposed explanations range from enlarged 
gauge structures and discrete flavor symmetries to more abstract algebraic or topological mechanisms. 
For example, in the $SU(3)_C\times SU(3)_L\times U(1)_X$ class of models, anomaly cancellation can 
correlate the number of fermion families with the number of colors \cite{Frampton1992,CorianoFrampton2024}. 
Recent work has also explored genuinely nonstandard mechanisms in which triplication follows 
from algebraic consistency conditions rather than from an explicit threefold copy imposed 
at the outset \cite{Patrascu2026}.

Here we ask a different and deliberately more modest set of questions:
\begin{quote}
Can a genuine three-body holonomy, together with the cyclic ordering structure, 
generate three family-like mass branches in a relativistic solvable model? 
Can it produce a hierarchy among the branches, 
and can the complete three-state loop leave an observable low-energy remnant 
when only the lightest state is directly accessible?
\end{quote}

A first motivation comes from exactly solvable relativistic few-body Dirac models 
with pairwise contact interactions \cite{Munakata1990,Sakamoto1993,SakamotoNakanoYoshikawa1993}. 
These models provide a controlled reference system in which the pairwise sector 
can be separated cleanly from an additional genuine three-body structure.
In our previous work \cite{Yoshikawa2026}, we introduced a genuine
three-body holonomy in this relativistic Dirac framework and showed
that, when implemented as global matching data around the
triple-coincidence point, it can enter the massive spectral sector.
The present work develops the phenomenological consequences of that
construction. In particular, we ask whether the six ordering sectors
and the associated cyclic structure can produce a three-branch mass
spectrum and whether the heavy branches can leave a low-energy
remnant.

A second motivation comes from topologically nontrivial three-body contact interactions 
in one dimension. For three nonidentical particles on a line, the triple-coincidence locus 
has codimension two in configuration space. After removing the center of mass, the coincidence 
becomes a puncture of the relative plane, and the complement admits nontrivial winding. 
The corresponding contact interaction can be represented by a twisted boundary condition 
or equivalently by a singular background gauge flux in configuration space \cite{Ohya2024}. 
This is closely analogous at the level of global phase information to the Aharonov--Bohm mechanism, 
where a gauge potential can produce a physically observable loop phase even 
when the local field vanishes along the particle path \cite{AharonovBohm1959}.

Our purpose is not to identify this low-dimensional configuration-space topology directly 
with the flavor structure of the SM. Rather, we use the one-dimensional relativistic three-body problem 
as a proof of concept. Three logically distinct ingredients must be kept separate throughout:
\begin{enumerate}
\item the punctured relative configuration space permits a continuous three-body loop phase $\theta_3$;
\item the six particle-ordering sectors, at fixed intrinsic parity, form a three-dimensional cyclic 
space with exactly three $C_3$ eigenchannels;
\item a Hermitian effective mass operator converts the loop phase into a three-branch spectrum.
\end{enumerate}
The topology therefore permits the phase but does not by itself determine the number of branches. 
The discrete triplication originates from the cyclic ordering algebra. Likewise, the low-energy 
elimination performed later requires a general three-state mass matrix after symmetry breaking; 
it is not obtained by arbitrarily eliminating two components of an exactly symmetric circulant matrix.

The central phenomenological logic established below is
\begin{equation}
\begin{aligned}
\text{topology} &+ \text{fixed-parity }C_3\text{ ordering}\\
&\longrightarrow \text{three branches},\\
\text{holonomy} &+ \text{cyclic coupling}\\
&\longrightarrow \text{mass splitting and hierarchy},\\
C_3\text{ breaking} &+ \text{heavy-state elimination}\\
&\longrightarrow \text{infrared loop remnant}.
\end{aligned}
\label{eq:chain}
\end{equation}

\section{Relativistic three-body Dirac model}

As a solvable relativistic reference system, we employ the one-dimensional few-body Dirac model 
developed by Munakata, Sakamoto, Ino, 
and collaborators \cite{Munakata1990,Sakamoto1993,SakamotoNakanoYoshikawa1993}. 
We consider three equal-mass Dirac particles on a line,
\begin{equation}
H_0=\sum_{i=1}^{3}\left(-i\alpha_i\partial_{x_i}+m\beta_i\right),
\label{eq:H0}
\end{equation}
Here the Dirac matrices are 
\begin{align}
 \alpha_i = \begin{pmatrix} 0&1\\1&0 \end{pmatrix}, \qquad 
    \beta_i=\begin{pmatrix} 1&0\\0&-1 \end{pmatrix}
\end{align}
which satisfy the standard relations $\alpha_i^2=\beta_i^2=1$ and $\{\alpha_i,\beta_i\}=0$ for each particle. 
Operators acting on different particle factors commute.

For the pairwise interaction we take the Sakamoto--Munakata--Ino contact form
\begin{equation}
V_2=g\sum_{i<j}(1-\alpha_i\alpha_j)\delta(x_i-x_j).
\label{eq:V2}
\end{equation}
Our purpose is not to modify the known pairwise sector but to ask whether an additional genuine 
three-body coincidence structure can support a nontrivial global phase that is absent 
from the pairwise matching alone. Schematically,
\begin{equation}
H=H_0+V_2+V_3,
\label{eq:Hfull}
\end{equation}
where $V_3$ denotes a genuine three-body interaction or, more generally, 
a self-adjoint boundary condition associated with the triple-coincidence locus
\begin{equation}
x_1=x_2=x_3.
\end{equation}

After removing the center of mass, introduce Jacobi coordinates
\begin{equation}
\rho=\frac{x_1-x_2}{\sqrt{2}},\qquad
\lambda=\frac{x_1+x_2-2x_3}{\sqrt{6}}.
\label{eq:jacobi}
\end{equation}
The triple coincidence is the origin $(\rho,\lambda)=(0,0)$ of the relative plane.

\section{Topological specialness of the three-body coincidence}

For $N$ particles on a line, removal of the center of mass leaves an $(N-1)$-dimensional relative space. 
If only the full $N$-body coincidence is removed, the relevant local topology is
\begin{equation}
\mathcal C_N^*\simeq \mathbb R^{N-1}\setminus\{0\}.
\end{equation}
For $N=3$,
\begin{equation}
\pi_1(\mathbb R^2\setminus\{0\})\simeq\mathbb Z,
\label{eq:pi1three}
\end{equation}
whereas
\begin{equation}
\pi_1(\mathbb R^d\setminus\{0\})=0,\qquad d\ge 3.
\end{equation}
Thus the full three-body coincidence is uniquely situated to support an ordinary winding holonomy 
in one spatial dimension. For general many-body configuration spaces with all triple coincidences removed, 
richer pure-twin-group structures occur \cite{Ohya2024}; our narrower statement concerns 
the puncture generated by the full three-body coincidence itself.

\begin{figure}[t]
\centering
\includegraphics[width=0.82\linewidth]{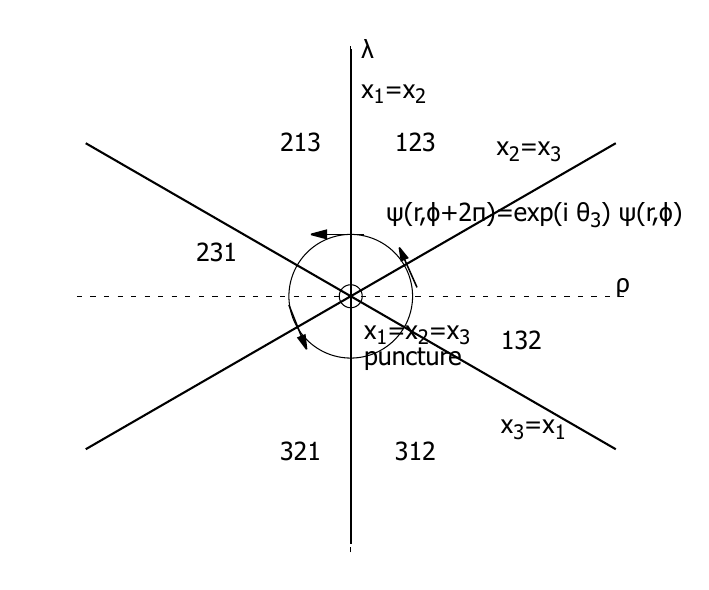}
\caption{Relative configuration space $(\rho,\lambda)$ after removal of the center-of-mass coordinate. 
The three pair-coincidence lines divide the plane into six ordering sectors. The triple-coincidence 
point is removed, producing a puncture that permits 
the winding boundary condition $\Psi(r,\phi+2\pi)=e^{i\theta_3}\Psi(r,\phi)$. With $\rho=(x_1-x_2)/\sqrt2$, 
the right half-plane has $x_1>x_2$, so the ordering labels shown are fixed accordingly.}
\label{fig:configspace}
\end{figure}

Equation \eqref{eq:pi1three} permits a twisted boundary condition
\begin{equation}
\Psi(r,\phi+2\pi)=\mathcal W_3\Psi(r,\phi),
\label{eq:twistedBC}
\end{equation}
where $\mathcal W_3$ is unitary. In a one-dimensional irreducible representation of 
the winding group one may write simply
\begin{equation}
\mathcal W_3=e^{i\theta_3},\qquad \theta_3\equiv\theta_3+2\pi.
\label{eq:scalarholonomy}
\end{equation}
The quantity $\theta_3$ is the gauge-invariant loop phase that will enter the effective three-channel 
theory below.

\section{Pairwise matching and trivial net pair holonomy}

The three-particle configuration space is divided into six ordering sectors,
\begin{equation}
123,\ 213,\ 231,\ 321,\ 312,\ 132.
\label{eq:sixsectors}
\end{equation}
Across an $ij$ coincidence line the pairwise contact condition may be written with the matching operator
\begin{equation}
J_{ij}(g)=\exp\!\left[-\frac{ig}{2}(\alpha_i-\alpha_j)\right].
\label{eq:Jij}
\end{equation}

For a symmetric bound-state ansatz we assign the momenta
\begin{equation}
(p_L,p_M,p_R)=(-i\kappa,0,+i\kappa)
\label{eq:momentumassignment}
\end{equation}
to the particles occupying the left, middle, and right positions. For a one-particle Dirac spinor we use
\begin{equation}
u(p)=\begin{pmatrix}1\\ p/(\epsilon_p+m)\end{pmatrix},
\qquad \epsilon_p^2=m^2+p^2.
\end{equation}
For $p=\pm i\kappa$, define
\begin{equation}
u_- =\begin{pmatrix}1\\-i\eta\end{pmatrix},\quad
u_0 =\begin{pmatrix}1\\0\end{pmatrix},\quad
u_+=\begin{pmatrix}1\\+i\eta\end{pmatrix},
\end{equation}
where 
\begin{equation}
\eta\equiv\frac{\kappa}{\epsilon+m}.
\end{equation}
In sector $123$ the spinor factor is
\begin{equation}
\chi_{123}=u_-\otimes u_0\otimes u_+,
\end{equation}
whereas in sector $213$ it is
\begin{equation}
\chi_{213}=u_0\otimes u_-\otimes u_+.
\end{equation}
Writing
\begin{equation}
c=\cos\frac g2,\qquad s=\sin\frac g2,
\end{equation}
one finds
\begin{equation}
J_{12}(g)=(c-is\alpha_1)(c+is\alpha_2).
\end{equation}
The matching condition requires
\begin{equation}
c\eta+s=0,
\end{equation}
so that
\begin{equation}
\eta=-\tan\frac g2.
\end{equation}
Using $\epsilon^2=m^2-\kappa^2$ gives, for the corresponding sign convention,
\begin{equation}
\epsilon=m\cos g,\qquad \kappa=-m\sin g,
\end{equation}
and the symmetric reference energy
\begin{equation}
E_3^{(0)}(g)=m+2m\cos g.
\label{eq:E30}
\end{equation}

Because the $\alpha_i$ acting on different particle factors commute, all generators $\alpha_i-\alpha_j$ commute. The net oriented product of the pairwise matching matrices around a closed circuit encircling the triple-coincidence point therefore cancels between crossings with opposite orientation,
\begin{equation}
U_{\rm pair}=I.
\label{eq:Upair}
\end{equation}
Thus a nontrivial winding phase cannot be generated solely by composing the known pairwise contact conditions. A genuine three-body boundary condition or equivalent configuration-space flux is required.

\section{Genuine three-body holonomy}

The topological information is encoded by Eq.~\eqref{eq:scalarholonomy}. If one wishes to retain 
an explicit relativistic internal generator, a convenient Hermitian involution is
\begin{equation}
Q_3=\alpha_1\alpha_2\alpha_3,\qquad Q_3^2=I.
\label{eq:Q3}
\end{equation}
Since the $\alpha_i$ are Hermitian and commute across particle factors, $Q_3$ is Hermitian. 
The unitary operator
\begin{equation}
\mathcal W_3^{\rm spin}(\theta_3)=e^{i\theta_3 Q_3}
\end{equation}
acts on a $Q_3$ eigenstate, $Q_3\ket q=q\ket q$ with $q=\pm1$, as
\begin{equation}
\mathcal W_3^{\rm spin}(\theta_3)\ket q=e^{iq\theta_3}\ket q.
\end{equation}
Equivalently, the angular quantum number in that internal sector is shifted as
\begin{equation}
\nu_q=n+\frac{q\theta_3}{2\pi},\qquad n\in\mathbb Z.
\end{equation}
This two-valued $q$ label is an internal Dirac-sector label and is \emph{not} the origin of 
the three branches derived below. The triplication arises independently from the cyclic structure 
of particle ordering. In the remainder of the effective three-channel discussion we work 
in a fixed internal sector and denote the resulting gauge-invariant loop phase by $\theta_3$.

\section{Parity decomposition of the six ordering sectors}

Group the six ordering states into two cyclic triplets,
\begin{align}
\ket{e_0}&=\ket{123},& \ket{e_1}&=\ket{231},& \ket{e_2}&=\ket{312},\\
\ket{f_0}&=\ket{321},& \ket{f_1}&=\ket{132},& \ket{f_2}&=\ket{213}.
\end{align}
The ordering-sector Hilbert space is
\begin{equation}
\mathcal H_{\rm ord}=\mathcal H_e\oplus\mathcal H_f,
\qquad \dim\mathcal H_{\rm ord}=6.
\end{equation}
Let $C$ denote cyclic relabeling,
\begin{equation}
1\to2,\qquad 2\to3,\qquad 3\to1,
\end{equation}
so that
\begin{equation}
C\ket{e_j}=\ket{e_{j+1}},\qquad C\ket{f_j}=\ket{f_{j+1}},\qquad C^3=I,
\end{equation}
with indices understood modulo three.

Spatial inversion about the center of mass reverses the left-to-right ordering,
\begin{equation}
P\ket{e_j}=\ket{f_j},\qquad P\ket{f_j}=\ket{e_j},
\end{equation}
so that
\begin{equation}
P^2=I,\qquad [P,C]=0.
\end{equation}
For parity eigenvalue $p=\pm1$, define
\begin{equation}
\ket{O_j^{(p)}}=\frac{1}{\sqrt2}(\ket{e_j}+p\ket{f_j}),\qquad j=0,1,2.
\label{eq:paritybasis}
\end{equation}
These satisfy
\begin{equation}
P\ket{O_j^{(p)}}=p\ket{O_j^{(p)}}.
\end{equation}
Thus parity decomposes the original six-dimensional ordering space into two invariant 
three-dimensional subspaces,
\begin{equation}
\mathcal H_{\rm ord}=\mathcal H_+^{(3)}\oplus\mathcal H_-^{(3)}.
\end{equation}
If the full Hamiltonian is parity invariant, $[H,P]=0$, a bound-state multiplet may 
be chosen with fixed intrinsic parity. The relevant ordering space is then
\begin{equation}
\mathcal H_p=\mathrm{span}\{\ket{O_0^{(p)}},\ket{O_1^{(p)}},\ket{O_2^{(p)}}\},
\qquad \dim\mathcal H_p=3.
\label{eq:Hp}
\end{equation}
This does not identify parity-related ordering states or discard half the original Hilbert space; 
it is the standard decomposition into invariant parity sectors.

\begin{figure}[t]
\centering
\includegraphics[width=0.88\linewidth]{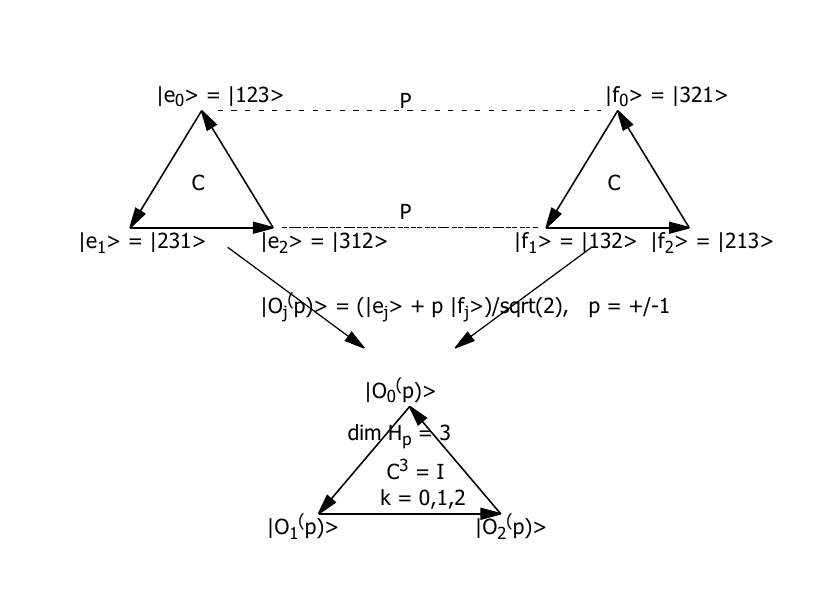}
\caption{Parity decomposition of the six ordering sectors. The two cyclic triplets $\{|e_j\rangle\}$ 
and $\{|f_j\rangle\}$ are exchanged by parity. At fixed intrinsic parity, the ordering space 
becomes three-dimensional, and the cyclic operator satisfies $C^3=I$, yielding exactly three 
eigenchannels $k=0,1,2$.}
\label{fig:triplication}
\end{figure}

\section{Exactly three cyclic eigenchannels}

Because $[P,C]=0$, the cyclic relabeling operator preserves each fixed-parity subspace. 
Diagonalizing $C$ within $\mathcal H_p$ gives the discrete Fourier modes
\begin{equation}
\ket{C_k^{(p)}}=\frac{1}{\sqrt3}\sum_{j=0}^{2}\omega^{-kj}\ket{O_j^{(p)}},
\qquad \omega=e^{2\pi i/3},\qquad k=0,1,2.
\label{eq:Cstates}
\end{equation}
They obey
\begin{equation}
C\ket{C_k^{(p)}}=\omega^k\ket{C_k^{(p)}}.
\end{equation}
The characteristic equation is
\begin{equation}
\det(\lambda I-C|_{\mathcal H_p})=\lambda^3-1=0,
\end{equation}
so the only cyclic eigenvalues are
\begin{equation}
1,\qquad \omega,\qquad \omega^2.
\end{equation}

At this point the distinction between topology and cyclicity is crucial. 
The nontrivial fundamental group in Eq.~\eqref{eq:pi1three} permits 
a continuous $U(1)$ loop phase $e^{i\theta_3}$. The three-valued label $k$ instead comes from 
the finite-dimensional ordering algebra. We therefore do \emph{not} identify the holonomy 
with an operator $e^{i\theta_3 C}$; because $C$ is unitary but not Hermitian, 
such an exponential is not in general unitary and would not yield phases $e^{i(\theta_3+2\pi k/3)}$. 
The correct combination of the two structures is realized through 
the gauge-invariant phase around the links of the Hermitian three-channel mass operator constructed next.

\section{Effective mass matrix and Peierls-type realization}

In the basis $\{\ket{O_0^{(p)}},\ket{O_1^{(p)}},\ket{O_2^{(p)}}\}$, the cyclic shift is represented by
\begin{equation}
C=\begin{pmatrix}
0&0&1\\
1&0&0\\
0&1&0
\end{pmatrix},\qquad C^3=I,\qquad C^\dagger=C^{-1}.
\label{eq:Cmatrix}
\end{equation}
Let $M$ be a Hermitian effective mass matrix in this fixed-parity space. Exact cyclic invariance requires
\begin{equation}
[M,C]=0.
\end{equation}
The most general Hermitian solution is a circulant matrix,
\begin{equation}
M_{\rm eff}=aI+bC+b^*C^{-1},\qquad a\in\mathbb R.
\label{eq:circulant}
\end{equation}
Equivalently, this form follows by projecting the full Hamiltonian onto the ordering subspace,
\begin{equation}
(M_{\rm eff})_{ij}=\mel{O_i^{(p)}}{H}{O_j^{(p)}},
\end{equation}
provided the microscopic dynamics preserves the cyclic symmetry.

\subsection{Peierls implementation of the loop phase}

The three ordering states may be viewed as three sites of an effective cyclic ring. 
A global holonomy is then represented in the standard Peierls manner: individual link phases 
depend on basis convention, whereas the phase accumulated around the complete loop 
is gauge invariant \cite{Peierls1933,AharonovBohm1959}.

Write the consistently oriented link amplitudes as
\begin{equation}
t_{10}=\Delta e^{i\phi_{10}},\qquad
t_{21}=\Delta e^{i\phi_{21}},\qquad
t_{02}=\Delta e^{i\phi_{02}},
\end{equation}
with
\begin{equation}
\phi_{10}+\phi_{21}+\phi_{02}=\theta_3\quad(\mathrm{mod}\ 2\pi).
\label{eq:loopphase}
\end{equation}
Under basis rephasing,
\begin{equation}
\ket{O_j^{(p)}}\to e^{i\chi_j}\ket{O_j^{(p)}},
\end{equation}
the link phases transform as
\begin{equation}
\phi_{ij}\to\phi_{ij}+\chi_i-\chi_j,
\end{equation}
while the sum in Eq.~\eqref{eq:loopphase} is unchanged. Thus only the total loop phase is physical.

A convenient symmetric gauge is
\begin{equation}
\phi_{10}=\phi_{21}=\phi_{02}=\frac{\theta_3}{3}.
\end{equation}
The Hermitian effective mass operator becomes
\begin{equation}
M_{\rm eff}=M_0I+\Delta e^{i\theta_3/3}C+\Delta e^{-i\theta_3/3}C^{-1}.
\label{eq:MeffPeierls}
\end{equation}
Explicitly,
\begin{equation}
M_{\rm eff}=\begin{pmatrix}
M_0&\Delta e^{-i\theta_3/3}&\Delta e^{+i\theta_3/3}\\
\Delta e^{+i\theta_3/3}&M_0&\Delta e^{-i\theta_3/3}\\
\Delta e^{-i\theta_3/3}&\Delta e^{+i\theta_3/3}&M_0
\end{pmatrix}.
\label{eq:Meffmatrix}
\end{equation}
The product of the three forward-link phases is
\begin{equation}
e^{i\theta_3/3}e^{i\theta_3/3}e^{i\theta_3/3}=e^{i\theta_3},
\end{equation}
as required. Equation \eqref{eq:MeffPeierls} is the minimal symmetric effective realization 
of the genuine three-body loop phase. It does not yet constitute a microscopic derivation 
of the magnitude $\Delta$ from a specific triple-coincidence operator; 
deriving $M_0$ and $\Delta$ from a self-adjoint microscopic three-body condition remains 
a central open problem.

\subsection{Three mass branches}

Using $C\ket{C_k^{(p)}}=\omega^k\ket{C_k^{(p)}}$, Eq.~\eqref{eq:MeffPeierls} gives
\begin{align}
M_{\rm eff}\ket{C_k^{(p)}}
&=\left[M_0+\Delta e^{i\theta_3/3}\omega^k+\Delta e^{-i\theta_3/3}\omega^{-k}\right]\ket{C_k^{(p)}}\\
&=M^{[k]}\ket{C_k^{(p)}},
\end{align}
where
\begin{equation}
M^{[k]}=M_0+2\Delta\cos\!\left(\frac{\theta_3+2\pi k}{3}\right),
\qquad k=0,1,2.
\label{eq:Mk}
\end{equation}
Thus the same continuous loop phase $\theta_3$ is sampled by three discrete cyclic channels.

\begin{figure}[t]
\centering
\includegraphics[width=0.78\linewidth]{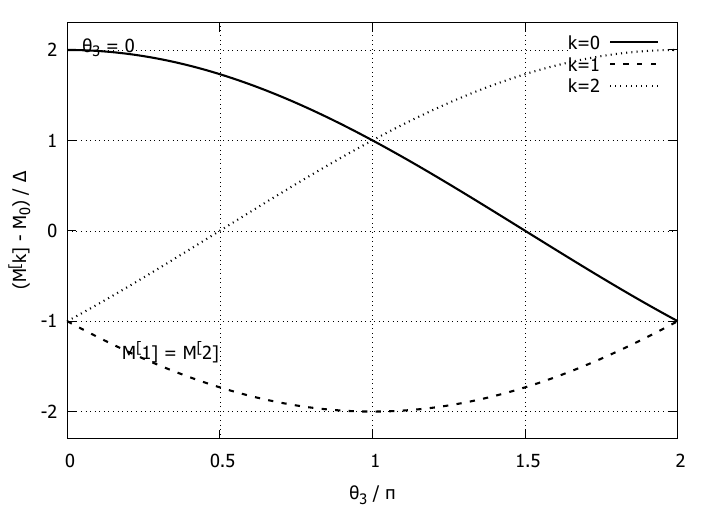}
\caption{Three mass branches as functions of the three-body holonomy angle. 
The plotted quantity is $(M^{[k]}-M_0)/\Delta=2\cos[(\theta_3+2\pi k)/3]$. At $\theta_3=0$, 
the conjugate pair $k=1,2$ is degenerate; a nonzero holonomy lifts this degeneracy.}
\label{fig:branches}
\end{figure}

For vanishing holonomy,
\begin{equation}
\theta_3=0,
\end{equation}
the spectrum is
\begin{equation}
M^{[0]}=M_0+2\Delta,\qquad
M^{[1]}=M^{[2]}=M_0-\Delta,
\end{equation}
so the conjugate $k=1,2$ channels are degenerate. For generic nonzero $\theta_3$ 
this degeneracy is lifted. At small phase,
\begin{equation}
M^{[1]}-M^{[2]}=-2\sqrt3\,\Delta\sin\!\left(\frac{\theta_3}{3}\right)
\simeq-\frac{2\sqrt3}{3}\Delta\theta_3.
\end{equation}

It is useful to introduce the Hermitian orientation-even and orientation-odd combinations
\begin{equation}
K_3=\frac{C+C^{-1}}{2},\qquad
J_3=\frac{C-C^{-1}}{2i}.
\end{equation}
Then
\begin{equation}
M_{\rm eff}=M_0I+2\Delta\cos\!\left(\frac{\theta_3}{3}\right)K_3
-2\Delta\sin\!\left(\frac{\theta_3}{3}\right)J_3.
\label{eq:KJ}
\end{equation}
The $J_3$ term changes sign under reversal of the cyclic orientation $C\leftrightarrow C^{-1}$. 
This orientation sensitivity is not automatically CP violation; a CP-odd interpretation requires 
the transformation of the full relativistic theory, including charge conjugation 
and any additional interactions.

\section{Phenomenological implications}

\subsection{Holonomy-induced hierarchy}

Equation \eqref{eq:Mk} does more than count three channels. It permits a hierarchy 
through destructive interference between the common mass $M_0$ and the holonomy-dependent cyclic 
contribution. A transparent point is
\begin{equation}
\theta_3=\pi,
\end{equation}
for which
\begin{equation}
M^{[0]}=M_0+\Delta,\qquad M^{[1]}=M_0-2\Delta,\qquad M^{[2]}=M_0+\Delta.
\end{equation}
If
\begin{equation}
M_0=2\Delta+\mu,\qquad |\mu|\ll|\Delta|,
\end{equation}
then
\begin{equation}
M^{[1]}=\mu,\qquad M^{[0]}=M^{[2]}=3\Delta+\mu.
\label{eq:hierarchy}
\end{equation}
One branch can therefore be parametrically lighter than the other two. This is a 
demonstrated cancellation mechanism within the effective theory, not yet an explanation 
of why the microscopic parameters should satisfy $M_0\simeq2\Delta$.

Near this point, let $\theta_3=\pi+\varepsilon$. Expanding Eq.~\eqref{eq:Mk},
\begin{align}
M^{[1]}&=M_0-2\Delta+\frac{\Delta}{9}\varepsilon^2+O(\varepsilon^4),\\
M^{[0]}&=M_0+\Delta-\frac{\sqrt3\Delta}{3}\varepsilon+O(\varepsilon^2),\\
M^{[2]}&=M_0+\Delta+\frac{\sqrt3\Delta}{3}\varepsilon+O(\varepsilon^2).
\end{align}
The light branch is stationary to first order, whereas the heavier pair splits linearly.

\subsection{Low-energy remnant of the complete three-state loop}

We next ask whether the two heavy states become completely irrelevant 
when the available energy is below their production thresholds. 
To formulate this question consistently, we allow diagonal splittings 
and other $C_3$-breaking effects and consider the general Hermitian matrix
\begin{equation}
\mathcal M=\begin{pmatrix}
m_1&t_{12}&t_{13}\\
t_{12}^*&M_2&t_{23}\\
t_{13}^*&t_{23}^*&M_3
\end{pmatrix},
\qquad |m_1|\ll |M_2|,|M_3|.
\label{eq:generalM}
\end{equation}
Here state 1 denotes the light mode after symmetry breaking, whereas states 2 and 3 are heavy. 
Equation \eqref{eq:generalM} need not retain the exact circulant symmetry of Eq.~\eqref{eq:Meffmatrix}; 
it is the appropriate infrared parameterization once a light-heavy basis has been dynamically selected.

Write
\begin{equation}
V=(t_{12},t_{13}),\qquad
M_H=\begin{pmatrix}M_2&t_{23}\\t_{23}^*&M_3\end{pmatrix}.
\end{equation}
Block elimination gives the exact eigenvalue condition
\begin{equation}
\lambda=m_1-V(M_H-\lambda I)^{-1}V^\dagger.
\label{eq:SchurExact}
\end{equation}
This is the standard Schur-complement or projection-operator structure familiar 
from effective Hamiltonian methods \cite{Lowdin1951,Feshbach1958}. 
For $|\lambda|\ll|M_{2,3}|$, the leading local low-energy mass is
\begin{equation}
M_{\rm low}=m_1-VM_H^{-1}V^\dagger
+O\!\left(\frac{\lambda |V|^2}{M_H^2}\right).
\label{eq:Mlowgeneral}
\end{equation}
Since
\begin{equation}
M_H^{-1}=\frac{1}{M_2M_3-|t_{23}|^2}
\begin{pmatrix}
M_3&-t_{23}\\
-t_{23}^*&M_2
\end{pmatrix},
\end{equation}
we obtain
\begin{equation}
M_{\rm low}=m_1-
\frac{M_3|t_{12}|^2+M_2|t_{13}|^2-2\,\mathrm{Re}(t_{12}t_{23}t_{13}^*)}
{M_2M_3-|t_{23}|^2}.
\label{eq:Mlowexactleading}
\end{equation}

Under independent rephasings of the three basis states, the closed product
\begin{equation}
\mathcal P_{123}=t_{12}t_{23}t_{31},\qquad t_{31}=t_{13}^*,
\end{equation}
is invariant. Define its phase by
\begin{equation}
\Theta_{123}=\arg(t_{12}t_{23}t_{31}).
\label{eq:ThetaLoop}
\end{equation}
When the effective three-state loop descends from the same oriented Peierls structure 
as Eq.~\eqref{eq:loopphase}, we identify
\begin{equation}
\Theta_{123}=\pm\theta_3
\end{equation}
up to the convention for loop orientation. The genuinely three-link contribution 
to Eq.~\eqref{eq:Mlowexactleading} is then
\begin{equation}
\Delta M_{\rm hol}=
\frac{2|t_{12}t_{23}t_{31}|}{M_2M_3-|t_{23}|^2}\cos\theta_3.
\label{eq:DeltaMhol}
\end{equation}

\begin{figure}[t]
\centering
\includegraphics[width=0.82\linewidth]{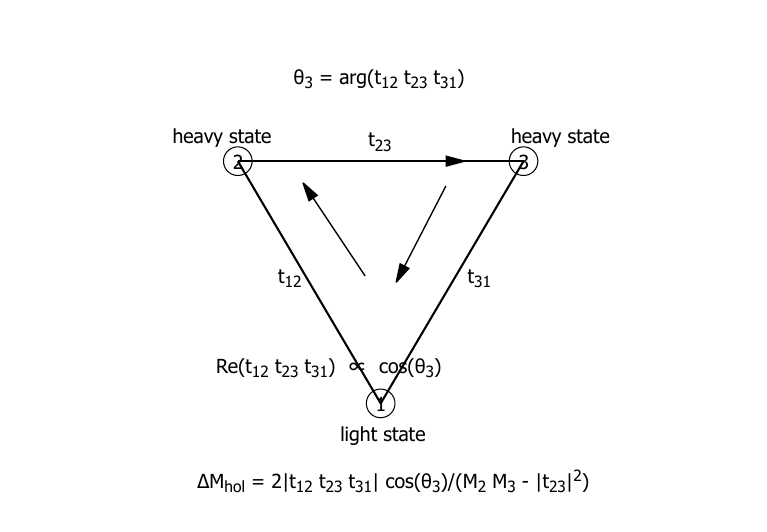}
\caption{Schematic three-state loop underlying the low-energy remnant. 
States 2 and 3 may be heavy and integrated out, while the light-state effective mass 
retains the rephasing-invariant three-link contribution 
proportional to $\mathrm{Re}(t_{12}t_{23}t_{31})\propto\cos\theta_3$.}
\label{fig:lowenergy}
\end{figure}

It vanishes if any one of the three links is absent. In the weak-mixing limit $|t_{23}|^2\ll|M_2M_3|$,
\begin{equation}
M_{\rm low}\simeq
m_1-\frac{|t_{12}|^2}{M_2}-\frac{|t_{13}|^2}{M_3}
+\frac{2|t_{12}t_{23}t_{31}|}{M_2M_3}\cos\theta_3.
\label{eq:Mlowweak}
\end{equation}
The first two terms are ordinary two-state mixing corrections. 
The last term requires the complete loop $1\to2\to3\to1$ and first appears at cubic order 
in the link amplitudes. Consequently, the heavy states decouple for fixed link amplitudes 
as $M_{2,3}\to\infty$, but for finite heavy masses they leave a phase-dependent infrared memory.

For the symmetric Peierls assignment of Eq.~\eqref{eq:MeffPeierls}, the product 
of three consistently oriented links is
\begin{equation}
t_{12}t_{23}t_{31}=\Delta^3 e^{\pm i\theta_3},
\end{equation}
so the same Wilson-loop phase that splits the three branches controls the genuinely 
three-link term in the low-energy mass.

\subsection{CP-sensitive phase: possibility and limitation}

Complex conjugation reverses the loop orientation,
\begin{equation}
\theta_3\to-\theta_3.
\end{equation}
The mass correction in Eq.~\eqref{eq:DeltaMhol} is even under this operation because
\begin{equation}
\mathrm{Re}(t_{12}t_{23}t_{31})=|t_{12}t_{23}t_{31}|\cos\theta_3.
\end{equation}
It therefore does not by itself constitute CP violation. The complementary 
orientation-odd rephasing invariant is
\begin{equation}
I_{\rm loop}=\mathrm{Im}(t_{12}t_{23}t_{31})
=|t_{12}t_{23}t_{31}|\sin\theta_3.
\label{eq:Iloop}
\end{equation}
This quantity supplies a necessary phase-sensitive ingredient for a CP-odd effect in an embedding 
where the basis and interactions give the loop an operational meaning. 
However, a single Hermitian mass matrix is not sufficient to predict a measurable CP asymmetry. 
As in familiar flavor physics, a physical CP-odd invariant requires relative information 
between noncommuting structures rather than an isolated phase convention \cite{Jarlskog1985}. 
A realistic application would therefore require at least one additional noncommuting mass matrix 
or interaction, together with an explicit charge-conjugation transformation and, for rate asymmetries, 
interfering amplitudes carrying the required absorptive structure.

\section{Toy phenomenology of holonomy-induced family triplication}

The purpose of this section is to separate two logically distinct ingredients: 
the existence of three cyclic branches and the splitting of their masses.  
The former follows from the fixed-parity ordering algebra, whereas the latter requires 
an effective interaction acting within that three-dimensional space.

\subsection{Triplication precedes mass splitting}

Suppose all physical quantum numbers other than the cyclic label are fixed, 
including the intrinsic parity $p$ of the bound-state multiplet.  
The parity label is therefore common to the three states and is not interpreted 
as a family index.  The toy identification is
\begin{equation}
\text{family-like index}\quad\longleftrightarrow\quad k\in\mathbb Z_3
\end{equation}
with $k=0,1,2$ defined by the three eigenvalues of $C$.

The origin of the triplication is therefore
\begin{align}
\text{six ordering sectors} + \text{fixed parity}
&\Rightarrow \dim\mathcal H_p=3,\\
C^3=I
&\Rightarrow k=0,1,2.
\end{align}
The continuous holonomy angle $\theta_3$ does not create the number three.  
It enters only after the three-dimensional cyclic space has been identified.

Before the effective cyclic coupling is switched on, the three branches may be viewed 
as a degenerate triplet with common reference mass $M_0$,
\begin{equation}
\Delta=0
\quad\Longrightarrow\quad
M^{[0]}=M^{[1]}=M^{[2]}=M_0.
\label{eq:degtriplet}
\end{equation}
Here $M_0$ denotes the common reference mass, while the superscript $[k]$ labels the cyclic branch.  
This notation avoids confusing the reference subscript in $M_0$ with the branch label $k=0$.

\subsection{Interaction-induced lifting of the degeneracy}

Turning on the minimal Hermitian cyclic interaction gives Eq.~\eqref{eq:MeffPeierls} and hence
\begin{equation}
M^{[k]}
=M_0+2\Delta\cos\!\left(\frac{\theta_3+2\pi k}{3}\right),
\qquad k=0,1,2.
\label{eq:Mkfamily}
\end{equation}
Thus the cyclic algebra provides the three branches, while the effective interaction assigns 
different masses to them.  At $\theta_3=0$,
\begin{equation}
M^{[0]}=M_0+2\Delta,
\qquad
M^{[1]}=M^{[2]}=M_0-\Delta,
\end{equation}
so the interaction first separates the $k=0$ branch from the conjugate pair.  
A generic nonzero $\theta_3$ then lifts the remaining $k=1,2$ degeneracy.  
In this sense the model contains a two-step structure:
\begin{equation}
\text{cyclic ordering}\Rightarrow\text{triplication},\\
(\Delta,\theta_3)\Rightarrow\text{mass splitting}.
\end{equation}

\subsection{Low-energy visibility of the full triplet}

A hierarchical spectrum does not imply that the heavy branches are irrelevant.  
If additional dynamics breaks exact $C_3$ symmetry and selects a light-heavy basis, 
the heavy states can contribute virtually.  The Schur-complement result in Eq.~\eqref{eq:Mlowweak} contains 
the genuinely three-link term
\begin{equation}
\Delta M_{\rm hol}
\propto
\mathrm{Re}(t_{12}t_{23}t_{31})
=|t_{12}t_{23}t_{31}|\cos\theta_3,
\end{equation}
which vanishes if any one of the three links is removed.  Consequently, 
even when only one light branch is kinematically accessible, 
the low-energy effective mass can retain information about the complete three-state loop.

This observation is phenomenologically useful because it separates direct production from virtual sensitivity.  It does not mean that the heavy family-like branches fail to decouple: for fixed link amplitudes the correction vanishes as their masses are taken to infinity.

\subsection{Scope of the family interpretation}

The identification of $k=0,1,2$ with a family-like quantum number is 
a phenomenological interpretation, not a derivation of the three Standard Model generations. 
An ordering-sector label in a few-body problem is not automatically a fundamental flavor index.  
The more limited result established here is that genuine three-body topology, 
a fixed-parity cyclic ordering algebra, and a holonomy-dependent effective interaction 
together provide a minimal mechanism with exactly three internal branches, controllable mass splitting, 
and a possible infrared remnant of the complete triplet.

A realistic family model would additionally have to show 
that the three branches carry identical Standard Model gauge quantum numbers, 
explain the observed mass hierarchy without parameter tuning, and generate quark and lepton mixing.  
The present construction is therefore complementary to gauge-anomaly, discrete-flavor, 
and algebraic approaches to family replication \cite{Frampton1992,CorianoFrampton2024,Patrascu2026}.

\section{Possible extension toward \texorpdfstring{$3+1$}{3+1} dimensions}

The literal configuration-space winding used above is special to low spatial dimension. 
In $3+1$ dimensions, removing a single full-coincidence point from the higher-dimensional 
relative configuration space does not generate the same ordinary winding group. Any continuation 
of the present mechanism must therefore reinterpret the $C_3$ structure as an \emph{internal} 
three-state holonomy rather than as the direct winding of point particles around a puncture.

Let $S$ be a unitary internal cyclic-shift operator,
\begin{equation}
S^3=I,\qquad S^\dagger=S^{-1},\qquad S\ket{k}=\omega^k\ket{k}.
\end{equation}
A Hermitian three-body interaction carrying an internal loop phase may then be written schematically as
\begin{equation}
V_3^{(3+1)}=F(\bm\rho,\bm\lambda)
\left[\Delta_3 e^{i\delta}S+\Delta_3 e^{-i\delta}S^\dagger\right],
\label{eq:V3_31}
\end{equation}
with real $\Delta_3$ and real scalar profile $F$. Equation \eqref{eq:V3_31} is manifestly 
Hermitian and has the same algebraic structure as Eq.~\eqref{eq:MeffPeierls}. In a genuine gauge 
or field-theoretic completion, the physically meaningful phase would again be 
a rephasing-invariant closed-loop quantity rather than an arbitrary phase assigned 
to a single matrix element.

This proposal should be regarded only as an algebraic continuation of the solvable 
one-dimensional prototype. A realistic $3+1$ dimensional embedding would have 
to explain (i) the microscopic origin of the internal three-state space, 
(ii) why three states carry identical $SU(3)_C\times SU(2)_L\times U(1)_Y$ quantum numbers 
before symmetry breaking, (iii) how chiral interactions act in that space, 
(iv) how quark and lepton mixing arise, and 
(v) how the effective link amplitudes and their loop phase descend from a local field theory. 
These questions are beyond the present model.

\section{Discussion}

The main value of the construction is that it separates several statements 
that can otherwise be conflated.

\subsection{What is derived}

Three results follow directly from the structure of the model.  
First, removing the full three-body coincidence from the relative plane gives $\pi_1\simeq\mathbb Z$, 
so a continuous $U(1)$ winding phase is topologically allowed.  
Second, the known Sakamoto--Munakata--Ino pairwise matching matrices 
have a trivial net oriented product around the triple-coincidence point; 
the nontrivial loop phase therefore cannot be attributed to the pairwise contacts alone.  
Third, at fixed intrinsic parity the six ordering sectors reduce 
to a three-dimensional invariant cyclic space with eigenvalues $1,\omega,\omega^2$.  
These statements establish the coexistence of a continuous three-body loop phase 
and an exactly three-valued cyclic label, but they do not by themselves determine a mass spectrum.

The pairwise matching used here is especially simple 
because the generators $\alpha_i-\alpha_j$ commute.  Consequently, consistency 
of successive pairwise crossings is automatically compatible 
with the corresponding factorized ordering changes.  
In that restricted sense no additional nontrivial Yang--Baxter constraint 
is needed to produce the threefold structure: the triplication comes 
from the fixed-parity $C_3$ ordering algebra, not from a nontrivial Yang--Baxter solution.  
In more general integrable systems, however, Yang--Baxter consistency 
provides the compatibility condition for different factorizations of many-body scattering 
or transfer processes~\cite{Baxter1972,Zamolodchikov1979}.  A genuinely noncommuting extension 
of the matching operators could therefore make Yang--Baxter consistency dynamically important, 
but that lies beyond the present model.

\subsection{What is an effective-model assumption}

The Peierls form
\begin{equation}
M_{\rm eff}=M_0I+\Delta e^{i\theta_3/3}C+\Delta e^{-i\theta_3/3}C^{-1}
\end{equation}
is the minimal Hermitian $C_3$-invariant realization that assigns 
the gauge-invariant loop phase $\theta_3$ to the three-channel ring.  
Its eigenvalues, Eq.~\eqref{eq:Mk}, are therefore exact within this effective three-state model.  
What has not yet been derived microscopically is the magnitude $\Delta$, the reference mass $M_0$, 
or the precise map from a self-adjoint triple-coincidence condition to these parameters.

Likewise, the low-energy Schur-complement analysis is not obtained by deleting two components 
of the exactly circulant matrix.  It applies only after additional dynamics breaks exact cyclic 
symmetry and selects one light state and two heavy states.  Once such a basis exists, 
the rephasing-invariant product $t_{12}t_{23}t_{31}$ provides a clean measure 
of the complete three-link loop and produces the infrared term proportional to $\cos\theta_3$.

\subsection{Hierarchy and infrared memory}

The spectrum can support a parametrically light branch through cancellation, 
for example near $\theta_3=\pi$ and $M_0\simeq2\Delta$.  This demonstrates 
that a hierarchy is possible, but not that it is natural.  Establishing naturalness requires 
a microscopic theory that fixes $M_0/\Delta$ or protects the cancellation by a symmetry.

The low-energy remnant is conceptually distinct from the hierarchy itself.  
Heavy branches may be absent from external states and yet contribute virtually through
\begin{equation}
\Delta M_{\rm hol}
\propto
|t_{12}t_{23}t_{31}|\cos\theta_3.
\end{equation}
The effect nevertheless decouples for infinite heavy masses at fixed mixing.  
Thus the model predicts finite-energy memory, not nondecoupling in the strict field-theoretic sense.

\subsection{CP and dimensional extension}

The mass spectrum and the low-energy mass correction are even in $\theta_3$ 
and therefore do not establish CP violation.  The orientation-odd invariant
\begin{equation}
I_{\rm loop}=\mathrm{Im}(t_{12}t_{23}t_{31})
\propto\sin\theta_3
\end{equation}
can become a CP-sensitive ingredient only when combined with additional 
noncommuting dynamics and an explicit definition of charge conjugation 
in the full relativistic theory \cite{Jarlskog1985}.

Finally, the real-space puncture responsible for the winding phase is special 
to the low-dimensional problem.  A $3+1$ dimensional continuation should therefore be interpreted 
as an internal three-state holonomy rather than as a literal dimensional lift 
of the same configuration-space winding.  The unresolved task is to identify 
a local field-theoretic origin for that internal space and its loop phase 
while preserving the chiral gauge structure of the Standard Model.

Taken together, these qualifications sharpen the status of the proposal: 
the topology and cyclic triplication are structural results; the Peierls mass operator 
is a minimal effective realization; the hierarchy and infrared memory are consequences 
of that realization; and the identification with fermion families remains 
a phenomenological hypothesis to be tested in a more microscopic theory.

\section{Conclusion}

We have constructed a relativistic one-dimensional three-body framework 
in which a genuine three-body loop phase and a three-valued cyclic ordering structure coexist 
but play different roles.  The triple-coincidence puncture permits a continuous $U(1)$ holonomy, 
while the six ordering sectors reduce at fixed intrinsic parity 
to a three-dimensional cyclic space with $k=0,1,2$.  The number 
of branches is therefore supplied by the ordering algebra rather than 
by the value of the holonomy angle.

A minimal Hermitian Peierls realization of the loop phase gives
\begin{equation}
M^{[k]}=M_0+2\Delta\cos\!\left(\frac{\theta_3+2\pi k}{3}\right),
\qquad k=0,1,2.
\end{equation}
In this notation $M_0$ is the common reference mass and $[k]$ labels the cyclic branch.  
For $\Delta=0$ the triplet is degenerate; turning on the cyclic interaction separates the branches, 
and a nonzero $\theta_3$ lifts the conjugate-channel degeneracy.  Near suitable parameter values, 
one branch can be parametrically lighter than the other two.

If additional dynamics subsequently breaks exact cyclic symmetry and defines 
a light-heavy basis, integrating out the heavy states yields 
a low-energy contribution proportional to the rephasing-invariant closed product
\begin{equation}
\Delta M_{\rm hol}
\propto
\mathrm{Re}(t_{12}t_{23}t_{31})
=|t_{12}t_{23}t_{31}|\cos\theta_3.
\end{equation}
This term requires the complete three-link loop and therefore provides a concrete infrared remnant 
of the full three-state structure.  Its orientation-odd partner, 
proportional to $\sin\theta_3$, is phase sensitive but is not by itself a CP-violating observable.

The construction should not be read as a derivation of the three Standard Model generations.  
It is a solvable phenomenological prototype showing that three-body topology, 
fixed-parity cyclic ordering, and a holonomy-dependent interaction can jointly produce exactly 
three family-like branches, a mass hierarchy, and a residual low-energy loop effect.  
The next decisive step is microscopic: derive $M_0$, $\Delta$, 
and the effective link amplitudes from a self-adjoint triple-coincidence condition 
or genuine three-body interaction, identify the mechanism that breaks exact $C_3$ symmetry, 
and determine whether an internal-space analogue can be embedded consistently 
in a $3+1$ dimensional chiral field theory.

\end{document}